\documentclass[f]{ceurart}
\usepackage[utf8]{inputenc}
\usepackage{amssymb}
\usepackage{pifont}
\usepackage{enumitem}
\usepackage{listings}
\usepackage{tabularx}
\usepackage{xcolor}

\newcommand{\xmark}{\ding{53}}%
\begin{document}
\copyrightyear{2024}
\copyrightclause{Copyright for this paper by its authors.
  Use permitted under Creative Commons License Attribution 4.0
  International (CC BY 4.0).}
\conference{Irish Conference on Artificial Intelligence and Cognitive Science (AICS), December 2024, Ireland}
\title{Datasheets for Healthcare AI: A Framework for Transparency and Bias Mitigation}
\author{Marjia Siddik}[email=marjia.siddik3@mail.dcu.ie]
\author{Harshvardhan J. Pandit}[email=me@harshp.com]
\address{ADAPT Centre, School of Computing, Dublin City University, Dublin, Ireland}

\begin{abstract}
The use of AI in healthcare has the potential to improve patient care, optimize clinical workflows, and enhance decision-making. However, bias, data incompleteness, and inaccuracies in training datasets can lead to unfair outcomes and amplify existing disparities. This research investigates the current state of dataset documentation practices, focusing on their ability to address these challenges and support ethical AI development. We identify shortcomings in existing documentation methods, which limit the recognition and mitigation of bias, incompleteness, and other issues in datasets. We propose the ‘Healthcare AI Datasheet’ to address these gaps, a dataset documentation framework that promotes transparency and ensures alignment with regulatory requirements. Additionally, we demonstrate how it can be expressed in a machine-readable format, facilitating its integration with datasets and enabling automated risk assessments. The findings emphasise the importance of dataset documentation in fostering responsible AI development.
\end{abstract}
\begin{keywords}
AI in Healthcare \sep Dataset Documentation \sep Data Transparency \sep Bias Mitigation \sep Ethical AI \sep GDPR \sep EU AI Act \sep Risk Assessment
\end{keywords}
\textcolor{red}{\small{Presented at 32nd Irish Conference on Artificial Intelligence and Cognitive Science (AICS) 2024}}
\maketitle

\section{Introduction}

AI has the potential to enhance diagnostics, treatment planning, patient monitoring, and overall care delivery \cite{topol2019high}. By utilising medical records and other data collected over time, AI can make healthcare more efficient, personalized, and accessible \cite{shah2019artificial}. However, its deployment also introduces ethical and legal challenges, particularly due to the use of sensitive data and the risk of harms. Prominent amongst known issues are biases \cite{obermeyer2016predicting} which exacerbate existing health disparities and create unequal treatment outcomes. Such biases can arise from training datasets, algorithmic practices, and incorrect deployments, and act to compromise the effectiveness, thereby threatening patient safety and undermining trust in healthcare systems \cite{obermeyer2019dissecting}. As a result, the use of AI poses challenges to the fairness and integrity of healthcare delivery, particularly when it relies on unrepresentative datasets \cite{obermeyer2019dissecting}, flawed data collection \cite{char2018implementing}, and reflects existing societal prejudices \cite{mehrabi2021survey}. These issues can lead to AI models that perform poorly for certain demographic groups, amplifying health disparities and compromising patient care \cite{benjamin2019assessing}. 

Beyond bias, other concerns include data incompleteness \cite{vayena2018machine}, inaccuracies \cite{rajkomar2018ensuring}, and outdated information \cite{jiang2017artificial}, all of which can undermine the effectiveness and reliability of AI systems. To assess whether such issues exist in the development and use of AI, it is essential to have comprehensive documentation regarding the origins, composition, limitations, and other contexts for how the AI system was developed - in particular the data used to train it \cite{gebru2021datasheets}. Without this information, AI systems cannot be reliably assessed for suitability of use, and can inadvertently cause harm or fail to deliver effective care \cite{paullada2021data}.

With data and AI enriched healthcare poised to significantly progress in the next decade based on legal advancements such as the EU Health Data Space regulation, it is important for uses of data and AI in healthcare to adhere to existing regulatory frameworks such as General Data Protection Regulation (GDPR) \cite{gdpr} and the Artificial Intelligence Act (AI Act) \cite{aiact}. Which means that dataset documentation practices should also incorporate information to support compliance with GDPR and AI Act so that issues such as bias and potential harms can be evaluated and enforced through legal mechanisms. At the same time, healthare is highly dependant on local contexts, where different regions and countries have differing frameworks and legislations for how the data gets generated and utilised in health research. Ireland recently published its Health Information Bill \cite{healthBill2024} in 2023 which enables the reuse of data for healthcare research. In such cases, it is also vital to evaluate whether dataset documentation practices are sufficient to support such initiatives.

This study therefore investigates the question: \textit{"How can dataset documentation support mitigation of bias and promote ethical AI in healthcare systems?"} and explores the answer through the following objectives:
\setlist{nolistsep}
\begin{enumerate}[label=\textbf{RO\arabic*},noitemsep]
    \item Identify categories of bias which should be documented (Section 2.1).
    \item Identify legal considerations which should be documented for datasets (Section 2.2).
    \item Evaluate existing dataset documentation practices regarding representation of identified bias categories and legal considerations (Section 2.3).
    \item Develop a machine-readable dataset documentation method that incorporates identified requirements and fills in gaps in current practices (Section 3).
    \item Discuss how the solution will work within the Irish healthcare context (Section 4).
\end{enumerate}

\section{Literature Review}

\subsection{Categorisation of Bias}
Bias in AI refers to the tendency of AI systems to produce results that reflect societal inequities, which is especially concerning in healthcare, where biased AI tools can have serious consequences. AI models trained on non-diverse datasets often fail to generalize across populations, potentially worsening healthcare disparities. For example, Celi et al. \cite{celi2022sources} observed that many AI datasets come from high-income countries like the US and China, reducing their relevance in low- and middle-income countries with distinct healthcare challenges. Despite the importance of mitigating these biases, existing dataset documentation practices often overlook these disparities, revealing major shortcomings in current frameworks \cite{russo2024leveraging}.

Several types of bias impact healthcare AI, leading to inequitable outcomes. \textit{Sample bias} occurs when training data does not adequately represent the target population, resulting in skewed predictions. Annotator bias emerges when individuals labeling the data introduce their prejudices, further distorting AI outputs. \textit{Temporal bias} arises from changes in data patterns over time, impacting AI model relevance \cite{gaonkar2020ethical}. For example, \textit{gender bias} in diagnostic AI, such as chest X-ray interpretation, has shown to skew results based on biological differences that arise over time \cite{ganz2021assessing}. 
Unfortunately, current documentation rarely addresses these biases comprehensively, pointing to the need for improved strategies.

Biases such as \textit{data-driven} and \textit{algorithmic bias} also contribute to healthcare inequality. Data-driven bias occurs when certain demographics are over-represented, while algorithmic bias reinforces patterns favoring majority groups \cite{ganz2021assessing}. \textit{Human bias} introduced by researchers or clinicians adds further complexity to fairness \cite{ganz2021assessing}. If dataset documentation practices do not distinguish between these different kinds of bias, or only report on specific ones (e.g. gender bias) - then it risks creating a false sense of security by assuming biases have been identified and addressed. Further, by not incorporating the information required to identify and address such biases, the dataset documentation also limits its usefulness.

Based on these, we identify a gap in current practices that needs to be address by having dataset documentation \textit{distinguish between the different categories of bias and should record that aids in identifying and addressing them}. This addresses \textbf{RO1}.

\subsection{Legal Frameworks Governing Bias}
The development and deployment of data-based AI systems in healthcare requires strict adherence to laws to ensure fairness, accountability, and transparency. Laws such as GDPR and AI Act require conducting impact assessments to protect human rights and prevent harms based on a risk-based approach where certain data and technologies are considered as \textit{high-risk} based on their sensitivity and potential risks. If dataset documentation approaches do not incorporate such requirements, or do not provide sufficient information to support implementing them, it leads to legal uncertainties, risks, and makes assessing \textit{liability} difficult - which is a vital incentive to ensure safety and security in technology.

The GDPR, which regulates processing of personal data, establishes specific categories, which includes health, of data as being \textit{special} (Article 9) - meaning they are more sensitive and merit a higher degree of consideration in risk management (Article 32) and impact assessments (Article 35). GDPR also establishes accountability based on the role of `\textit{Controller}' where an entity determines the `means and purposes' of processing data, where processing covers any collection, storage, use, sharing, and erasure of personal data. Further, the GDPR also establishes rights (Articles 12-23) associated with data - such as the requirement to provide notices, ability to opt-out of automated decision making, right to be forgotten, rectification, and erasure. When datasets constitute personal data (as defined by GDPR Article 4), their collection and use, as well as potentially the AI systems developed using them are likely to be subject to the GDPR. Without sufficient information, users of data miss out on the \textit{safety net} provided by the GDPR in terms of safety and security obligations when reusing data, and end up creating complications and potential liabilities for themselves as they do not have documented evidence of the dataset's quality and GDPR compliance.

The AI Act, a recent development, establishes risk levels for use of AI, and has obligations regarding transparency, human oversight, and data management. Similar to the GDPR, documented evidence is vital for obligations under the AI Act to assess and demonstrate that data, or AI developed using data, is compliant with safety and reliability standards. More specific to Ireland, the Health Information Bill (2023) establishes the creation and sharing of digital health records and provides a framework for the reuse data for scientific research and public benefit. In this, it requires specific documented assessments of data and uses of AI similar to the obligations of GDPR to ensure appropriate practices regarding security and safety.

From this, we establish that dataset documentation practices should incorporate \textit{information to support legal obligations regarding transparency and accountability - most specifically the provenance and legal categorisation of data, risk and impact assessments, and involvement of entities in specific legal roles} This addresses \textbf{RO2}.

\subsection{Current Dataset Documentation Practices}\label{sota:documentation}

Comprehensive documentation practices in the machine learning community often receive limited attention beyond immediate technical information, and no standardised processes exist to ensure transparency, accountability, reproducibility, interoperability, and quality - especially in contexts such as healthcare where non-reporting of issues such as bias can cause harms \cite{gebru2021datasheets}. To address this, several proposals have been published, of which we focus on notable ones that are widely known or are in the scope of our work. 

`Datasheets for Datasets' \cite{gebru2021datasheets}is a seminal work that defines information requirements to record motivation, composition, and usage aspects to enhance transparency and reproducibility. While it acknowledges regulations such as GDPR and issues such as bias, it does not provide for recording specific risks such as different categories of biases and does not align its information with regulatory requirements. `Dataset Nutrition Label' \cite{chmielinski2022dataset}, developed by the Data Nutrition Project, aims to ``enhance context, contents, and legibility'' by ``providing at-a-glance information''. It contains information required for bias identification but does not address measures for  mitigation or specifying further risks or regulatory requirements.

`Open Datasheets' \cite{roman2023open, heger2022understanding} provide a machine-readable format designed to improve dataset discoverability and usability, but does not expand upon risk assessment and regulatory information. `Data Statements for NLP' \cite{bender2018data} enables recording information with the goal of supporting bias mitigations, but does not account for different risks or regulations. Tools such as DataDoc Analyzer \cite{giner2023datadoc} and MetaReader \cite{jannah2014metareader} support ensuring completeness and bias identification based on existing approaches, but do not explore expanding their information requirements.

These existing approaches\footnote{Due to spacial limitations, an overview of existing approaches is provided later as part of our proposed approach.} show a necessity to document information regarding datasets so as to inform and support the `data value chain' in addressing risks - such as biases - and avoiding harms. However, they have limitations in terms of acknowledging different risks beyond a few bias categories (e.g. gender or sex), do not support systematic risk assessments, and more critically are not aligned with regulatory requirements - which makes their enforcement and use in accountability difficult. In the context of healthcare, we could not find any specific approach which adapts or explores the specific collection and (re-)use of data in a clinical or medical research context. Further, healthcare settings typically have additional policies and guidelines established within the institution, consortium, or as sectorial regulations - which can only be supported in dataset documentation practices if they are extensible. Additionally, the information required to be documented can come from different entities - including inter-organisational units - which necessitates standardisation and interoperability to ensure its effectiveness. We could not find any approaches which tackled these aspects.

From this analysis, we determined that the use of data for AI in healthcare settings requires a solution that addresses existing gaps regarding expanded bias categorisations, risk assessments, and compliance with regulations. This addresses \textbf{RO3}.

\section{Developing an Improved Machine-Readable Datasheet}
Based on the analysis of the state of the art, we identified the need to create an improved dataset documentation approach that supports information requirements regarding bias categorisation and risk assessment, and is aligned with the GDPR (\textbf{RO4}). We also identified the need for structured machine-readable representations to support maintaining and providing datasheets alongside the data. For this, we selected the `Datasheets for Datasets' \cite{gebru2021datasheets} approach as a baseline given its existing prevalence and impact, and extended it to incorporate our additional requirements. Our proposed datasheet is 
\textbf{available online}\footnote{\url{https://github.com/marjiasdk/Healthcare-AI-Datasheet}} with an example schema.

\subsection{Methodology}
We first identified and analysed the information that could be recorded from using existing approaches for datasheet documentation and found three gaps (bias categories, risk assessment, regulations), for which we then developed specific requirements to document information. Through an iterative process, we developed a structured datasheet by starting with 18 identified information fields from the `Datasheets for Datasets' \cite{gebru2021datasheets} approach, and extended it to over 50 fields in the final iterations. The additional fields were developed based on requirements to document information associated with identified categories of bias - such as temporal characteristics and demographics, risk assessment information - such as provenance of data, completeness, existing or potential measures, and regulatory information - such as applicable laws and impact assessments. In addition to this, we also made explicit the information fields associated with purposes for which the dataset was created, and its intended uses, and usage restrictions - which can aid the process of determining suitable data reuses and avoid misuses. For existing fields, we focused on specificity where vague fields were refined for clarity (e.g. usage restrictions and data characteristics). 

In order to evaluate the effectiveness of developed information requirements, we sought to identify existing datasets on popular platforms such as Kaggle and Hugging Face whose documentation contained this information. We could not identify a suitable dataset as most datasets did not contain even the preliminary information required by existing dataset documentation practices, and going through their associated publications and reports would have required exorbitant amounts of time\footnote{The lack of findability mechanisms based on machine-readable data also contributed to these difficulties.}. Therefore, we undertook manual exercises to create documentation for hypothetical datasets based on different scenarios such that all information fields would be populated. Through this process, we identified several refinements based on ambiguity in information (e.g. date format), necessity to provide a controlled vocabulary (e.g. to express likelihood), and additional fields (e.g. usage prohibitions derived from risk assessments). 

We then developed a JSON based structure to represent the datasheet in a machine-readable format. We chose JSON as it is a popular data format that is natively supported in all major programming languages, is easily communicated on the web, and enables a structured schemas that can be validated for completeness, correctness, and compliance. We also chose JSON as it is easy to learn and iterate prototypes for a developing schema. For future interoperability and standardisation, we recommend using existing standards such as DCAT\footnote{\url{https://www.w3.org/TR/vocab-dcat/}} with ODRL\footnote{\url{https://www.w3.org/TR/odrl-model/}} for expressing usage policies and DPV \cite{pandit2024dataprivacyvocabularydpv} to represent regulatory information.

\subsection{Description of Information Fields}\label{datasheet:fields}
There are 55 information fields broadly categorised in 10 sections as follows. \\
\textbf{Metadata} These fields contain information describing the dataset in terms of its title, version, publisher, and license. 
\\
\textbf{Purpose}  These fields describe the purposes for which the dataset was created (e.g. clinical research), what was the intended benefit (e.g. improve diagnostic accuracy), and the intended beneficiaries (e.g. healthcare providers, patients).
\\
\textbf{Source Information} These fields describe the source and origin of data, and whether the collection process had (ethical) approval (e.g. from organisation) and its funding sources.
\\
\textbf{Temporal Information} These fields describe temporal aspects of the data in the dataset, such as which period it covers (e.g. 2019-2023) and last updated (e.g. December 2023).
\\
\textbf{Demographic Information} These fields describe demographic information for individuals whose data is present, such as age and age ranges (e.g. 18-65), gender, and ethnicity, as well as fields to indicate the \textit{likelihood} of specific kinds of bias due to the demographic distributions.
\\
\textbf{Data Characteristics} These fields describe the media type for data (e.g. images), and also indicate whether the data is \textit{incomplete} along with the missing elements and reasons.
\\
\textbf{Bias Mitigation Methods} These fields describe bias mitigation methods that have already been applied as well as suggested measures to adopters.
\\
\textbf{Personal Data} These fields describe whether the data constitutes as \textit{personal} (e.g. non-anonymized patient records), specific categories of personal data (e.g. name, age), its sensitity (e.g. low), and for (partially-)anonymised data - which anonymisation techniques were used and its risk of reidentification.
\\
\textbf{Risk and Compliance} These fields provide a way to indicate the risk levels (separately for generic and legal), jurisdiction and applicable laws (e.g. EU and GDPR), existence of impact assessments (e.g. a GDPR DPIA), and suggested mitigation measures (e.g. auditing security risks prior to data reuse).
\\
\textbf{Usage Restriction} These fields define limitations and constraints on the (re-)use of the dataset, such as through access restrictions (e.g. only use within organisation), specific permissions (e.g. only used for cancer research) or prohibitions (e.g. no third party sharing), and obligations (e.g. reciprocity to share results back with data provider).

\subsection{Comparison with Existing Approaches}

Table \ref{table:comparison} compares our developed Datasheet approach with existing established approaches from Section \ref{sota:documentation} - namely the Datasheets for Datasets \cite{gebru2021datasheets}, Dataset Nutrition Labels \cite{chmielinski2022dataset}, and Data Statements for NLP \cite{bender2018data}. Due to spatial limitations of this article, we only provide a summary overview of this comparison, with the full analysis available 
online\footnote{\url{https://github.com/marjiasdk/Healthcare-AI-Datasheet/blob/main/comparison-sota.csv}}. 
In the table, the rows represent the information in sections (as described in Section \ref{datasheet:fields}, and the values represent whether the information is present ($\bullet$), has some fields missing ($\circ$), or is not present (\xmark). The last two rows consider whether the approach requires structured information (e.g. a consistent vocabulary) and is interoperable by being machine-readable (e.g. using JSON).

\begin{table}[ht]
    \centering
    \caption{Comparative analysis of our proposed Datasheet with existing documentation approaches}
    \footnotesize
    \begin{tabularx}{\columnwidth}{|p{3.25cm}|X|X|X|X|}
    \hline
        \textbf{Category} & \textbf{This Approach} & \textbf{Datasheets for Datasets \cite{gebru2021datasheets}} & \textbf{Dataset Nutrition Label \cite{chmielinski2022dataset}} & \textbf{Data Statements for NLP \cite{bender2018data}} \\ \hline
        \textbf{Metadata} & $\bullet$ & $\bullet$ & $\bullet$ & $\bullet$ \\ \hline
        \textbf{Purpose} & $\bullet$ & $\bullet$ & $\bullet$ & $\bullet$ \\ \hline
        \textbf{Source Information} & $\bullet$ & $\bullet$ & $\bullet$ & $\circ$ \\ \hline
        \textbf{Temporal Information} & $\bullet$ & $\circ$ & $\circ$ & \xmark \\ \hline
        \textbf{Demographics} & $\bullet$ & \xmark & $\circ$ & $\bullet$ \\ \hline
        \textbf{Data Characteristics} & $\bullet$ & $\circ$ & $\circ$ & $\circ$ \\ \hline
        \textbf{Bias Mitigations} & $\bullet$& \xmark & \xmark & \xmark \\ \hline
        \textbf{Personal Data} & $\bullet$ & $\circ$ & $\circ$ & $\circ$ \\ \hline
        \textbf{Risk and Compliance} & $\bullet$ & $\circ$ & $\circ$ & \xmark \\ \hline
        \textbf{Usage Restriction} & $\bullet$ & $\circ$ & $\bullet$ & \xmark \\ \hline
        \textbf{Machine-readable} & $\bullet$ & \xmark & $\bullet$ & \xmark \\ \hline
        \textbf{Interoperability} & $\circ$ & \xmark & $\circ$ & \xmark \\ \hline
    \end{tabularx}
    \normalsize
    \label{table:comparison}
\end{table}

From this comparison, we see how the proposed advances the state of the art by highlighting important gaps in current approaches and providing our solution as the path forward. Most prominently, we address the crucial issue of missing information in dataset documentation practices regarding data characteristics and temporal information which is necessary to identify and mitigate commonly found biases. We also addressed risk and (legal) compliance more thoroughly which enables existing legal mechanisms and obligations to be used to support and enforce accountability and prevent harms that may arise from data collection and (re-)use. Finally, we also address the lack of providing documentation as structured information and ensuring it is interoperable and machine-readable. In this, our approach does not provide the best possible solution as it does not propose a standardised or standards-based representation of information - though it does show why this is required and how it can be achieved (e.g. using semantic web standards of DCAT, ODRL, and DPV as mentioned in Section 3.2) which are promising areas for future work.

Though not visible from the overview table, the legal and ethical risks documented in the extended datasheet set it apart from existing approaches in a crucial and important manner. For example, our approach explicitly considers potential legal risks for re-identification risks and data sensitivity which are essential considerations when dealing with sensitive healthcare data, especially under regulations like GDPR. These risks are often only briefly touched upon in existing frameworks but are critical to be documented as having been considered for ensuring legal and policy compliance. In healthcare contexts, the sensitivity of patient data combined with the severe consequences of non-compliance can result in harm to patients, privacy violations, and legal repercussions. Therefore, our datasheets specifically support the higher level of scrutiny required in health and biomedical research to ensure patient safety and regulatory compliance by providing information fields that must be documented alongside the dataset.

Another important distinguishing factor is the demographic information section which goes beyond simple demographic documentation by requiring both information of distributions reflected in the dataset and also how such distributions may introduce bias into AI models. We feel this is a more comprehensive and in-depth approach that is also pragmatic for the healthcare context as compared to frameworks like Datasheets for Datasets which do not assess potential demographic bias in this way. In our datasheets, as the fields for distributions are explicit, the onus of identifying potential risks starts from the creation of the dataset, and any missing value (e.g. gender distribution unknown) becomes a risk in itself. To address this, we also provide a mitigations field which can enable a data provider to recommend that before using the dataset, an assessment of the bias should be carried out to eliminate or reduce potential issues.

\section{Application in Irish Healthcare Context}
The Irish healthcare system has been slow to adopt universal healthcare and has struggled with inconsistent data management practices across hospitals and clinics, influenced by historical resistance from the Catholic Church and private medical practitioners \cite{wren2019european}. This has led to fragmented reforms and slow legislative and infrastructural progress, creating challenges in healthcare data infrastructure, and impacting the implementation of effective AI systems. Gaps in leadership and strategy in health information management have also slowed the development of a connected system, hindering the fairness and effectiveness of AI systems \cite{craig2018understanding}.

Despite initiatives like IHIs and ePrescribing, investment in health information systems and ICT remains low in Ireland \cite{walsh2021developments}. Efforts such as the DQI framework seek to improve data quality, ensuring it is complete and reliable for creating fair AI systems \cite{hickey2021data}. While improving infrastructure is pivotal for reliable healthcare data, unaddressed biases could lead to legal and ethical issues, such as discrimination or violations of regulations like the GDPR. The two-tier healthcare model further complicates universal data-sharing standards as each organsiation relies on its own methods and data management practices.

While measures like the Health Identifiers Act 2014 and the establishment of Health Information and Quality Authority (HIQA) aimed to improve health data governance, the absence of well-defined and regulated dataset documentation practices continues to be a challenge. Areas such as data governance, legal uncertainties, and the development of central platforms to support health data research remain underexplored, making it difficult to create a unified system for equitable healthcare access \cite{walsh2021developments}. The lack of centralised policies that mandate uniform data practices hinders the development of unified datasets, which are critical for the implementation of fair and effective healthcare AI systems.

The Health Information Bill \cite{healthBill2024} (HIB) aims to resolve some of these challenges by creating a legal obligation for the state to set up specific healthcare research infrastructures and facilitate the reuse of data for research. It is expected to tie in to the European Health Data Spaces regulation which is currently under the legislative process and is expected to be finalised in 2025. More prominently, the HIB addresses the current healthcare system’s lack of coordination and disconnect between data management processes among organisations. While not explicitly addressing AI, the HIB does provide a legal framework for the reuse of healthcare data for research purposes and in conjunction with the recently published AI Act will guide the use of AI in healthcare for the near future. 

Our developed datasheet provides a necessary and timely approach  to address both the technical and legal challenges by establishing a structured and uniform approach to dataset documentation that is aligned with legal requirements from GDPR and which will support the requirements of Irish and European regulations regarding risk and impact assessments. Through this work, we have shown that the current prevalent dataset documentation practices are not sufficient for the current legal landscaope in Ireland and also do not support the practicalities of intra-organisational interactions which require documented information to avoid uncertainties and liabilities. 

By promoting transparent, accountable, and bias-conscious dataset documentation, our datasheet can serve as an important tool in helping Ireland’s healthcare sector prepare for these regulations and cultivate a practice of risk assessment across the data and AI lifecycles. Additionally, it supports the existing documentation requirements in current and upcoming regulations as well as policy frameworks such as HIQA’s guidelines. The possibility of representing the information in machine-readable form also opens up opportunities to automate the processes associated with creating the datasheets up to date with changes in AI systems, and to automate risk assessments by using datasheets as inputs that are passed along to stakeholders downstream within the AI value chain. Therefore, we recommend further developing and requiring the use of datasheet documentation practices along with supporting infrastructure and policies based on our work to enable and promote the ethical and legal reuse of data using AI across the Irish healthcare system. This completes our (\textbf{RO5}).

\section{Conclusion \& Future Work}

Our research highlights the importance of comprehensive dataset documentation in mitigating biases and promoting ethical AI in healthcare. Existing frameworks, such as Datasheets for Datasets and Dataset Nutrition Labels, often lack a specific focus on bias mitigation, particularly in the healthcare context. To address these gaps, we developed the Healthcare AI Datasheet, which incorporates detailed demographic information, data collection methods, and explicit bias mitigation strategies. This approach enhances transparency, accountability, and fairness in AI development, ensuring that systems reflect diverse populations and contribute to reducing healthcare disparities. Additionally, the machine-readable version of the datasheet facilitates integration into AI workflows, promoting responsible and ethical practices. By thoroughly documenting dataset characteristics and potential biases, healthcare providers can make more informed decisions when deploying AI systems, ensuring more equitable care. This is especially vital in preventing biased datasets from leading to misdiagnoses or unequal treatment outcomes.

While the Healthcare AI Datasheet represents notable progress, it has limitations. Its primary focus on healthcare datasets may restrict its broader applicability to other domains. Furthermore, its effectiveness relies on the accuracy and completeness of the information provided by dataset creators, which can introduce variability. Future research should prioritize real-world testing across healthcare settings to validate its effectiveness and refine its components. Implementing the datasheet in ongoing AI projects will help assess its ability to mitigate bias and improve compliance with regulations such as GDPR and the EU AI Act. Expanding its evaluation to AI systems deployed in other sectors could reveal its broader applicability and effectiveness. Additionally, refining the machine-readable version for better integration with diverse AI systems and exploring its use beyond healthcare could further enhance its impact, providing a scalable solution for ethical AI development across industries.

For the Irish healthcare context, there is a push for implementing a national framework that enables the wider sharing and reuse of healthcare data - especially for secondary purposes - and which is aligned with the future implementation of European Health Data Spaces (EHDS). We believe our approach for creating datasheets based on GDPR will facilitate this approach, and therefore would benefit from its application and refinement through use in real-world use-cases such as in hospitals and other clinical research settings.

\begin{acknowledgments}
This work was funded by the Health Research Board (HRB) through the Summer Student Scholarships (SS) scheme awarded to Marjia Siddik.
The ADAPT SFI Centre for Digital Media Technology is funded by Science Foundation Ireland through the SFI Research Centres Programme and is co-funded under the European Regional Development Fund (ERDF) through Grant\#13/RC/2106\_P2.
\end{acknowledgments}

\bibliography{references}

\end{document}